\documentclass{kluwer}    
\usepackage{graphicx}

\begin{document}
\begin{article}
\begin{opening}
\title{3D Simulations of the Local Bubble}
\author{D. \surname{Breitschwerdt}}
\institute{Max-Planck-Institut f\"ur Extraterrestrische Physik,
        Giessenbachstra{\ss}e, Postfach 1312, 85741 Garching, Germany; Email: breitsch@mpe.mpg.de}
\author{M. A. \surname{de Avillez}}
\institute{Department of Mathematics, University of \'Evora, R.
Rom\~ao Ramalho 59, 7000 Evora,
Portugal; Email: mavillez@galaxy.lca.uevora.pt }

\runningauthor{Breitschwerdt and de Avillez}
\runningtitle{Simulations of the Local Bubble}

\begin{abstract}
  We have performed high resolution 3D simulations of the Local
  Bubble (with 1.25 pc finest resolution) in a
  \emph{realistic background ISM}, jointly with the
  dynamical evolution of
  the neighbouring Loop I superbubble. We can reproduce (i) the
  size of the bubbles (in contrast to similarity solutions),
  (ii) the interaction shell with Loop I,
  discovered with ROSAT, (iii) predict the merging of the two
  bubbles in about 3 Myr, when the interaction shell
  starts to fragment, and, (iv) the generation of blobs like the
  Local Cloud as a consequence of a dynamical instability.
\end{abstract}

\end{opening}

\section{Introduction}
The Local Bubble (LB) is a cavity, elongated towards the Galactic
North Pole, with an average extension of about 100 pc, copiously
radiating in soft X-rays. There exist a number of discrepancies
between observations and modelling that have all been outlined in
a recent panel discussion (Breitschwerdt \& Cox 2004).
Most models proceed from a multi-supernova origin of the LB (cf.\
Bergh\"ofer \& Breitschwerdt 2002 for possible supernova (SN)
progenitors), in which the LB is the result of successive SN
explosions, although part of the soft X-ray emission could be of
heliospheric origin, generated by charge exchange reactions
between solar wind ions and heliospheric plasma.

\section{Results and Discussion}
We use a 3D Godunov type parallelized AMR hydrocode to track small
scale structures down to 1.25 pc, where necessary (for details
see Avillez 2000). According to Bergh\"ofer \& Breitschwerdt
(2002) 20 stars of Pleiades subgroup B1 with masses between 10 and
20 ${\rm M}_\odot$ were moving through the LB within the last 14
Myr, exploding after their main sequence life time, $\tau_{\rm
ms}$, along a path crossing $x=175$, $y=400$ pc (see Fig.~1), thus generating the
Local Cavity into which the LB expands. The Galactic SN rate has
been used for the setup of {\em other SNe} in the background disk.
\begin{figure}[t]
  \centering
\vspace*{1.in}
Left panel: breitsch$\_$fig1a \hspace*{0.5cm} Right panel: breitsch$\_$fig1b
\vspace*{1.in}

\caption{\emph{Left:} Temperature map (Galactic plane cut) of a 3D
LB simulation at present (i.e.\ 14.4 Myr after first explosion)
with the LB centered at (175, 400) pc and Loop I 200 pc to the
right. \emph{Right:} Same, but at $t=17$ Myrs, showing
fragmentation of the interaction shell and formation of cloudlets
by hydrodynamic instabilities. \label{fig1}}
\end{figure}
First we derive analytic similarity solutions, taking an initial
mass function for Galactic OB associations with powerlaw index
$\Gamma=-1.1$, and using $\tau_{\rm ms} = 3\,10^7 \, (m/[10 {\rm
M}_\odot])^{-\alpha}$ yr (Stothers 1972), with $\alpha = 1.6$, for
stars within the mass range $7\, {\rm M}_\odot \leq m \leq 30 \,
{\rm M}_\odot$. The mechanical energy input rate (see Bergh\"ofer
\& Breitschwerdt 2002) then is $L_{\rm SB} = L_0 \,
t_7^{\delta}$,
where $L_0 = 4.085 \times 10^{37} \,{\rm erg/s}$, $\delta =
-(\Gamma/\alpha + 1) = -0.3125$ and $t_7 = t/10^7$ yr. Modifying
suitably the similarity solutions of McCray \& Kafatos (1987), we
obtain $R_b = A t^\mu = 251 \left(2 \times 10^{-24} {\rm g}/{\rm
cm}^3 /\rho_0\right)^{1/5} t_7^{\mu} \, {\rm pc}$ with $\mu =
(2-\Gamma/\alpha)/5=0.5375$. Note, that our value of $\mu$ is
between the canonical value of $0.4$ for a Sedov-type solution and
$0.6$ for a wind type solution, due to the declining energy input
rate with time. If we now try to match a present day average
radius of the LB of even 146 pc at time $t_{\rm dyn}=14.4$ Myr, we
need a constant ambient density of $n_0\simeq 40 \, {\rm
cm}^{-3}$, a value way above the average ISM density.
Comparison with our excellently matching numerical results shows,
however, that this discrepancy must be due to mass loading, and
turbulence in a SN disturbed \emph{background} medium.

\end{article}

\begin{thebibliography}{}
\bibitem{} Avillez, M.A.: 2000, {\it MNRAS} {\bf 315}, 479.
\bibitem{bb2002} Bergh\"ofer, T. and Breitschwerdt, D.: 2002, {\it A\&A} {\bf 390}, 299.
\bibitem{} Breitschwerdt, D. and Cox, D.P.: 2004, {\it ApSS}, in press (astro-ph/0401428).
\bibitem{} McCray, R., and Kafatos, M.: 1987, {\it ApJ} {\bf 317}, 190.
\bibitem{} Stothers, R.: 1972, {\it ApJ} {\bf 175}, 431.

\end{thebibliography}
\end{document}